\documentstyle[12pt,epsf]{article}

\newcounter{subseqn}
\newcounter{saveeqn}

\renewcommand{\thesection}%
 {\Roman{section}.\setcounter{equation}{0}}
\renewcommand{\thesubsection}%
 {\setcounter{subseqn}{\value{equation}}%
\thesection\alph{subsection}%
\setcounter{equation}{\value{subseqn}}}
\renewcommand{\theequation}%
 {\mbox{\arabic{section}.\arabic{equation}}}

\newcommand{\alpheqn}%
 {\setcounter{saveeqn}{\value{equation}}%
\stepcounter{saveeqn}\setcounter{equation}{0}%
\renewcommand{\theequation}%
 {\mbox{\arabic{section}.\arabic{saveeqn}\alph{equation}}}}

\def\gtrsim{\raisebox{-4pt}{$\stackrel{\textstyle >}{\sim}$}} 

\def\lesssim{\raisebox{-4pt}{$\stackrel{\textstyle <}{\sim}$}} 

\begin{document} 

\title{Dynamical System Analysis for Inflation with Dissipation} 

\author{H. P. de Oliveira\thanks{E-mail: oliveira@symbcomp.uerj.br} $\;$
and Rudnei O. Ramos\thanks{E-mail: rudnei@symbcomp.uerj.br} \\ 
{\it Universidade do Estado do Rio de Janeiro, }\\ 
{\it Instituto de F\'{\i}sica - Departamento de F\'{\i}sica Te\'orica,}\\ 
{\it 20550-013 Rio de Janeiro, RJ, Brazil}} 

\date{August 1997} 

\maketitle 

\thispagestyle{empty} 

\begin{abstract} 

\baselineskip 24pt 

We examine the solutions of the equations of motion for an expanding
Universe, taking into account the radiation of the inflaton field
energy. We then analyze the question of the generality of inflationary
solutions in this more general setting of a dissipative system. We find
a surprisingly rich behavior for the solutions of the dynamical system of
equations in the presence of dissipational effects. We also determine
that a value of dissipation as small as $\sim 10^{-7}\,H$ can lead to a
smooth exit from inflation to radiation. 

\vspace{0.34cm} 
\noindent 
PACS number(s): 98.80 Cq 

\end{abstract} 

\newpage 

\setcounter{page}{1} 

\section{Introduction} 

It has been by now around 16 years since the idea of inflation
\cite{revinflat} was introduced in order to solve some of the underlying
problems of the standard big bang theory. Since then, inflation has
become one of the most important paradigms of early Universe cosmology.
The main idea of inflation is the existence of a period of accelerated
expansion of the Universe, when its energy density becomes dominated by
a potential energy density $V(\phi)$ of a scalar field $\phi$ (the
inflaton). The dynamics of inflation depend on the specifics of the many
models available describing inflation, but the basic mechanisms are
based on the equation of motion of a (homogeneous) inflaton field,
represented by its classical equation of motion 

\begin{equation} 
\ddot{\phi} + 3 H \dot{\phi} + V'(\phi) = 0 \;,
\label{class} 
\end{equation} 

\noindent (overdot represent derivative with respect to time and $V' =
\frac{dV}{d \phi}$) together with those of gravity. 

The basic assumption of most inflationary models is the one related with
the so called slow-roll approximation: at some initial time $t_0$, the
scalar field has a value far from the minimum of the potential
$V(\phi)$. At this time the scalar field energy density dominates and
the Universe enters the inflationary era. Sufficient inflation, required
to solve the main problems of the standard big bang theory, demands that
$\phi$ rolls down to the minimum sufficiently slowly. It is a common
belief that during this stage the particle like matter components are
all rapidly red-shifted away (as well the temperature, if we initially
started in a thermal bath) and that the evolution of $\phi$ can be well
represented by (\ref{class}). Later, when $\phi$ hits the bottom of the
potential and starts oscillating in a time scale shorter than the Hubble
time, all the energy stored in the oscillating field $\phi$ is then
released in the form of light particles that $\phi$ may be coupled to.
These decay products then thermalize, reheating the Universe. This final
stage of inflation is the so called reheating stage [for an account of
new developments in the theory of reheating, see, for instance,
\cite{revreheat} and references therein].

More recently, however, Berera and Fang \cite{Fang} have shown that
these two distinct stages of inflation can be compounded if the inflaton
dissipates its energy into a thermal bath. They have shown that this new
scenario is still consistent, providing sufficient e-folding of
inflation and generation of density perturbations of the right
magnitude, and that these density perturbations are consequences of
thermal fluctuations, as opposite to quantum fluctuations, generated
during the slow-roll stage, in the supercooled inflationary scenario.
This new picture for inflation has been denominated {\it warm inflation}
\cite{berera1,berera2}. 

In this modified description of inflation, we must alter (\ref{class})
to take into account the continuous dissipation of the inflaton field in
a thermal bath. A phenomenological way of doing this is, for example, by
introducing a friction-like term $\eta(\phi) \dot{\phi}$ in
(\ref{class}). $\eta(\phi)$ is the dissipation coefficient that comes
from the damping or ``decay'' of the scalar field $\phi$ interacting
with an environment or bath degrees of freedom, constituted by, for
example, light fields or field modes that $\phi$ may be coupled to. The
form of $\eta(\phi)$ depends on the details of the interaction terms.
The inclusion of a friction term in the equation of motion of the
inflaton field, to study its evolution in the early stages of inflation,
is not new. It has long been recognized that dissipative processes,
associated with the inflaton decay during its evolution, are able to
slow down the rolling of the field $\phi$ and, therefore, that these
processes would be able to support inflation in inflationary models such
as new or chaotic inflation \cite{oldstudies}. Recent works on
microscopic approaches to nonequilibrium dynamics of quantum fields
\cite{berera2,rud,boya1} support the introduction of a friction term of
the above form in the field equation of motion, especially under certain
conditions, such as near-equilibrium and when a loop expansion is valid.
These studies (see also, \cite{hu,boya2}) have also demonstrated the
generality of dissipation (as well as fluctuations) in the dynamics of
fields in interaction with a bath (thermal or quantum). Thus, from the
moment that we consider that the inflaton is coupled to light fields,
which is a necessary condition for describing reheating, we must also
consider the possible effects of backreactions of these fields, which
include dissipation through the decay of the inflaton also during the
stage of inflation. 

In principle we may consider, among other things, that quantities such
as sufficient inflation, the magnitude of initial density perturbations,
the strength of coupling constants of the inflaton field with other
fields during the inflationary era, not only constrain, but also dictate
how important dissipation is during inflation. {}For instance it is
known \cite{brand} that small coupling constants are necessary in order
that the slow-roll method to be a valid approximation, as well as for
obtaining the right magnitude of initial density perturbations.
Equivalently, density perturbations constrain quantum fluctuations to
remain small during the slow-roll stage of inflation. On the other hand,
in the warm inflation picture of \cite{Fang,berera1,berera2}, thermal
fluctuations, related to quite strong inflaton dissipation during
slow-roll, have been shown to be compatible with generation of initial
density perturbations, with minimal requirements on the inflationary
potential, besides that of the slow-roll approximation. 

Recent developments in nonequilibrium dynamical processes in the early
Universe, in particular during reheating, and the warm inflation
picture, lead us to consider some basic questions concerning dissipation
during inflation. Not only a proper microphysical approach is necessary
to understand the appearance of a non-negligible inflaton dissipation at
the early stages and/or during inflation, but we also need to
investigate how the classical inflaton dynamic gets changed in a
dissipative environment. In this paper we will be concerned with the
latter question, since it can shed some light about how specific
prescriptions for dissipation may change the dynamics in a
pre-inflationary era, and then, possibly, help to constrain field
theoretic model candidates to describe inflation in this new setting. A
detailed microscopic model will be presented in a forthcoming paper. The
qualitative analysis of the resulting set of field equations is the
first step in understanding radiation energy properties of the Universe
and in discriminating how general inflation is in dissipative systems.
{}Such analysis includes the asymptotic behavior of the solutions near
the critical points representing the initial singularity and the basic
features of the trajectories in the phase space that will be important
for the choice of initial conditions used in the numerical experiments.
We perform this study in the framework of the theory of dynamical
systems \cite{bogoy}. We then generalize earlier works on this subject
[see the Refs. \cite{belins1,belins2}], which were previously done in
terms of the classical set of equations for gravity plus that of the
inflaton field $\phi$, without dissipation. This study is important both
to understand the dynamics of inflation in a scenario like warm
inflation, as it is also an example of a dissipative dynamical system,
which may have other important applications in early Universe cosmology.

The paper is organized as follows: In Sec. II we give the basic
equations, define the region in phase space of interest to our study and
discuss some of their properties for different kinds of dissipation. In
Sec. III we present an analysis of the dynamical system of equations in
the presence of dissipation and determine the relevant asymptotic
solutions of the system of equations. In Sec. IV we study how
dissipation changes the number of physically relevant (advantageous)
trajectories, i.e., those having, e.g., a long enough inflationary
stage. We also give limiting values of when dissipation is of relevance
in the inflationary scenario. Sec. V is devoted to discussions of our
results and conclusions. 

\section{The Dynamical System} 

We start by writing the basic equations that characterize our dynamical
system: 

\begin{equation} 
H^2 = \frac{8 \pi G}{3} (\rho_\phi + \rho_{\rm rad}) -
\frac{k}{a^2} \;, 
\label{G1} 
\end{equation} 

\begin{equation} 
2 \dot{H} + 3 H^2 + \frac{k}{a^2} = - 8 \pi G (p_\phi +
p_{\rm rad}) \;, 
\label{G2} 
\end{equation} 

\begin{equation} 
\ddot{\phi} + 3 H \dot{\phi} + V'(\phi) + \eta(\phi)
\dot{\phi} =0 \;, 
\label{field} 
\end{equation} 

\noindent where $H= \dot{a}/a$ is the Hubble parameter and $G= 1/M_{\rm
pl}^2$, with $M_{\rm pl}$ the Planck mass. $k=0,+1,-1$ for a flat,
closed or open Universe, respectively. $\rho_{\phi (\rm rad)}$ and
$p_{\phi (\rm rad)}$ are the energy density and pressure for $\phi$
(radiation), respectively. We also have the standard relations:
$\rho_\phi = \frac{1}{2} \dot{\phi}^2 + V(\phi)$, $p_\phi = \frac{1}{2}
\dot{\phi}^2 - V(\phi)$ and $p_{\rm rad} = \frac{1}{3} \rho_{\rm rad}$.
Equations (\ref{G1}) and (\ref{G2}) are the standard gravity equations
derived from the Einstein equations. Eq. (\ref{field}) is the equation
of motion for the (homogeneous) scalar inflaton field $\phi$, with
$V(\phi)$ the effective potential for $\phi$. We have introduced
dissipation through a friction term $\eta(\phi) \dot{\phi}$ in the
equation of motion.

Equation (\ref{field}) deserves some important considerations. First, in
a realistic model the field dependence of the friction term may not be
exactly like the one given in (\ref{field}), particularly in highly
non-equilibrium situations and in a time dependent background, as some
recent studies have indicated \cite{boya2}. It has been demonstrated
\cite{rud,boya1}, however, that $\eta(\phi)\dot{\phi}$ describes quite
well the dissipation of the field $\phi$ in a bath environment in
near-equilibrium and when a perturbative expansion is valid. In spite of
this and without a better understanding from a field theoretic point of
view of nonequilibrium systems , we take for simplicity the friction
term in the form given in (\ref{field}). Since our analysis is mostly
qualitative, we expect that we will not miss much of the important
behavior of the system given a different prescription for dissipation.
Secondly, we take $\eta(\phi)$ as a polynomial in $\phi$. This is the
case for standard polynomial interactions among fields, at least when a
loop expansion in the fields is valid. The field dependence of
$\eta(\phi)$ is highly dependent of the type of interaction we consider
of the scalar field $\phi$ with the fields (or modes) that describe the
bath environment. {}For example, $\eta(\phi)$ may in some approximations
be well represented by a constant in the case of linear couplings (like
$\phi\psi$) \cite{berera2}, or be quadratic in $\phi$ for quadratic
couplings (like $\phi^2 \psi^2$) \cite{rud}. $\eta(\phi)$ may also have
a nontrivial dependence on coupling constants and temperature, as is the
case at high temperatures, which is the situation of interest in warm
inflation scenarios. {}Further considerations on $\eta(\phi)$ will be
considered below. {}Finally, Eq. (\ref{field}) must be viewed as the
equation obtained by averaging over a noise field, which is directly
related to dissipation through some generalized fluctuation-dissipation
relation. This means we are considering the inflaton field as a
classical background \cite{boya1} and we are looking at the averaged
behavior of the inflaton field in time. This is a good approximation as
far as we look to the inflaton dynamics at a sufficiently long period of
time when compared to the typical ``wavelength'' of the fluctuations,
given by $\sim [V'']^{-1/2}$. Our scale of time will be basically the
duration of inflation, which is at least 20 times larger than that value
(for e-folds of inflation larger than 70), as we will see in Sec. IV.
Also, the amplitude of fluctuations, given by the noise field two-point
correlation function, should not be too large during the inflaton
evolution, implying, roughly, that our approximation of taking the
ensemble averaged equation of motion is better the smaller is the
dissipation. Keeping in mind these considerations, we proceed with the
definition of the equations representing our dynamical system. 

We have that $\rho_\phi$ and $\rho_{\rm rad}$ evolve in time
as \cite{revinflat}: 

\begin{equation} 
\dot{\rho}_{\phi} + 3 H \dot{\phi}^2 + \eta(\phi)
\dot{\phi}^2 =0 \;, 
\label{dphi} 
\end{equation} 

\begin{equation} 
\dot{\rho}_{\rm rad} + 4 H \rho_{\rm rad} - \eta(\phi)
\dot{\phi}^2 =0 \;. 
\label{drad} 
\end{equation} 

\noindent 
Assuming a flat Universe ($k=0$), from (\ref{G1}), we can
consider 

\begin{equation} 
\rho_{\rm rad} = \frac{3}{8 \pi G} H^2 - \rho_\phi =
\frac{3}{8 \pi G} H^2 - \frac{1}{2} \dot{\phi}^2 - V(\phi) 
\label{drad2}
\end{equation} 

\noindent 
as the first integral of (\ref{drad}). Using $p_\phi$ and
$p_{\rm rad}$ in (\ref{G2}), we can write for $\dot{H}$ the equation: 

\begin{equation} 
\dot{H} = -2 H^2 - \frac{8 \pi G}{6}\left( \dot{\phi}^2
- 4 V(\phi) \right) \;. 
\label{Hdot} 
\end{equation} 

\noindent 
The equations (\ref{field}), (\ref{Hdot}), together with
$\dot{\phi} = \frac{d \phi}{dt}$, define a three-dimensional
(dissipative) dynamical system in the phase space of
$(\phi,\dot{\phi},H)$. 

{}For definitiveness, we take the simplest potential $V(\phi)$, which
describes a massive scalar field $\phi$: $V(\phi)= \frac{1}{2} m^2
\phi^2$. The generalization of our results for more complicated
potentials, including self-interactions terms and symmetry breaking,
should not present too much difficulty. Next, we take $\eta(\phi)$, as
discussed later, in the form $\eta(\phi) = \eta_n \phi^n$. We can
further constrain the form of $\eta(\phi)$ by noting that the friction
coefficient must be positive definite due to entropy requirements and
the parity invariance of the potential. Thus, $\eta_n > 0$ and $n=0$ or
an even integer. We analyze two main cases of possible interest: that of
$n=0$, in which case $\eta$ reduces to a constant; and that of $n=2$,
when $\eta \propto \phi^2$. Substituting $V$ and $\eta$ in (\ref{field})
and (\ref{Hdot}) and redefining variables $t \to t/m$, $\phi \to
\frac{1}{\sqrt{8 \pi G}} x$, $\dot{\phi} \to \frac{m}{\sqrt{8 \pi G}}
y$, $H \to m z$, $ \eta_0 \to m \beta_0$ and $\eta_2 \to 8 \pi G m
\beta_2$, we obtain the dynamical system: 

\begin{eqnarray}
\dot{x} &=& y 
\nonumber \\
\dot{y} &=& -3 y z - x - \beta_n x^n y
\label{sys2} \\
\dot{z} &=& -2 z^2 - \frac{1}{6} y^2 + \frac{1}{3} x^2
\;. \nonumber
\end{eqnarray}

The physical region in the phase space $(x,y,z)$ is defined by the
condition $\rho_{\rm rad} \geq 0$ which, in the units of (\ref{sys2}),
corresponds to: 

\begin{equation} 
3 z^2 - \frac{1}{2} y^2 - \frac{1}{2} x^2 \geq 0 \;,
\label{phys} 
\end{equation} 

\noindent 
or, expressing the above relation for $z$, $z^2 \geq
\frac{1}{6} (x^2 + y^2)$, which defines a conical volume in $(x,y,z)$,
with the physical region $\rho_{\rm rad} \geq 0$ inside the cone.
Expanding universes $(H>0)$ correspond to those trajectories for which
$z>0$ (inside the upper half of the conical volume). {}For our simple
quadratic potential there is only one critical point, namely, the origin
that represents the Minkowski spacetime. In general it is an attractor
and the form of approach depends upon the dissipation term. In fact we
can distinguish two main types of approach depending on the kind of
dissipation we may have. If $n=0$, we can readily show from
(\ref{sys2}) that the trajectories on the physical region approach 
the origin in the direction given by the $z$-axis. The $z$-axis itself
represents the usual Friedmann model with pure radiation and is, in
this case, an attractor for the late stages of field evolution. If
$\beta_0<2$ the approach to the $z$-axis is oscillatory, while for
$\beta_0 \geq 2$ it is exponential. If $n=2$ the approach is always
oscillatory tending to the origin on the surface $\rho_{rad}=0$. In
Figs. 1(a) to 1(c) we plot the field trajectories by a direct numerical
integration of the system (\ref{sys2}) for some representative values of
initial conditions and dissipative coefficients. 

In the absence of dissipation, the conical surface $\rho_{rad}=0$ is an
{\it invariant manifold}. As a consequence, all initial conditions
chosen on this surface will produce trajectories that evolve on it and
end at the origin. In particular, the dynamical system for which
$\rho_{rad}=0$ has been analyzed by Belinskii et. al.
\cite{belins1,belins2}. The effect of dissipation is to destroy the
invariant manifold since 

\begin{equation} 
\dot{\rho}_{\rm rad}|_{_{\rho_{\rm rad}=0}} =
\eta(\phi) \dot{\phi}^2 \neq 0 \;. 
\label{noninv} 
\end{equation} 

\noindent 
In this case the trajectories may no longer lie completely in
the surface $\rho_{\rm rad}=0$, even if the initial conditions are taken
on it. An important question we can ask is: in which way do the curves
cross the surface $\rho_{\rm rad}=0$ ? A simple way to answer this
question is, for example, by taking the scalar product at $\rho_{\rm
rad}=0$ between the vector normal to the outside surface of the cone,
$\vec n \propto - \vec \nabla \rho_{\rm rad} = (x,y,-6z)$, with the
velocity field vector of the trajectories on phase space, $\vec v =
(\dot x, \dot y, \dot z)$: 

\begin{equation} 
\vec n . \vec v |_{\rho_{\rm rad}=0} \propto - \beta_n
x^n y^2 \;. 
\label{nv} 
\end{equation} 

\noindent 
Since we are considering $\beta_n > 0$ and $n$ even, we obtain
that $\vec n . \vec v |_{\rho_{\rm rad}=0} <0$, which means that those
trajectories that initially lie inside the cone ($\rho_{\rm rad} >0$
region) remain inside the cone. The only trajectories that can cross the
cone lie initially outside it. This is a consequence of the requirement
of $\eta(\phi)$ be always positive definite. In particular we note that
for an odd integer $n$, we may have (\ref{nv}) positive for $x<0$ and we
could have trajectories leaving the region $\rho_{\rm rad} > 0$ to
$\rho_{\rm rad} <0$, in violation of the second law of thermodynamics. 

The inflationary region in phase space is defined by 

\begin{equation} 
\frac{\ddot{a}}{a} = \dot{H} + H^2 >0 \;,
\label{aceler} 
\end{equation} 

\noindent 
which, in terms of $x$, $y$ and $z$, gives 

\begin{equation} 
-z^2 -\frac{1}{6} y^2 +\frac{1}{3} x^2 > 0 \;.
\label{inflatreg} 
\end{equation} 

\noindent 
Trajectories satisfying (\ref{inflatreg}) are the ones that
are in the inflationary regime. The intersection between the regions
defined by (\ref{phys}) and (\ref{inflatreg}) defines the inflationary
physical region of interest to us. 

\section{Analysis at Infinity and Asymptotic Solutions} 

The analysis at infinity provide us with important information
concerning the behavior of solutions in the region where $x^2+y^2+z^2
\rightarrow \infty$. Such a region contains information about the early
stages of the universe just after the quantum domain, and therefore can
be described by the classical equations. An important issue to be
discussed is whether inflation plays the role of an attractor for most of
solutions no matter the presence of dissipation. {}For our work, the
following coordinate transformation is found to be useful: 

\begin{equation} 
x=\frac{u}{w},\,\, y=\frac{v}{w},\,\, z=\frac{1}{w},
\label{spher} 
\end{equation} 

\noindent 
where infinity is represented by the invariant manifold $w
= 0$. The projections of the physical region $\rho_{rad} \geq 0$ as well
as the inflationary domain $\frac{\ddot{a}}{a} > 0$ on this plane,
yield, respectively, the regions described as $u^2 + v^2 \leq 6$ and
$2\,u^2 - v^2 \geq 0$. The idea underlying the choice of coordinates
(\ref{spher}) is to compactify infinity, so that a dynamical system
in these variables describes properly the basic features of the
asymptotic behavior of the solutions. In Fig. 2, we present a better
view of the compactified region given by the cap of the conical region. 

The dynamical system (\ref{sys2}) in the coordinates $u, v, w$ is 

\begin{eqnarray} 
\frac{d\,u}{d\,\tau} &=&
u\,w^{\frac{n}{2}}\,\left( 2 + \frac{v^2}{6} - \frac{u^2}{3}\right) +
v\,w^{\frac{n}{2} + 1} \nonumber \\
\frac{d\,v}{d\,\tau} &=& v\,w^{\frac{n}{2}}\,\left( -1 +
\frac{v^2}{6} - \frac{u^2}{3} \right) - u\,w^{\frac{n}{2} + 1} -
\beta_{n}\,u^n\,v\,w^{1 - \frac{n}{2}} 
\label{sysinf} \\ 
\frac{d\,w}{d\,\tau} &=&
w^{\frac{n}{2} + 1}\,\left( 2 + \frac{v^2}{6} - \frac{u^2}{3}\right) \;,
\nonumber
\end{eqnarray} 

\noindent where we have changed the time parameter in such a way that
$d\,t = w^{\frac{n}{2} + 1}\,d\,\tau$, which is valid for $n = 0, 2$.
The critical points at infinity lie on the plane $w=0$ and their nature
depend on the type of dissipation we are considering. They represent the
initial singularity since, from Eqs. (3.5)-(3.8) below, we see that the
scale factor goes to zero at these points. In addition, at the critical
points the scalar of curvature $R=6\,(\dot{H} +2\,H^2)=
\frac{m^2}{w^2}\,(2\,u^2-v^2)$ diverges. As stated before two cases will
be analyzed, namely, $n = 0$, for constant dissipation, and $n=2$, for
field dependent dissipation. 

\subsection{Case n=0} 

We adopt the following steps here and for the case $n=2$ to analyze the
region at infinity $x^2+y^2+z^2 \rightarrow \infty$. First, we determine
all critical points on the surface $w = 0$ that are located in the
region defined by $0 \leq \theta \leq \theta^*$ (cf. Fig. 2) ($\tan
\theta^* = \sqrt{6}$), or in terms of the new coordinates (\ref{spher}),
inside the region $u^2 + v^2 \leq 6$. Secondly, in performing the
stability analysis, we may obtain approximate solutions near those
critical points and lying on the physical region. Thirdly, the knowledge
of unphysical trajectories that lie entirely on the surface $w = 0$ is
important since, by continuity arguments, such trajectories determine the
behavior of the integral curves of the system that lie within the cone
but near its surface \cite{belins2}. 

After substituting $n=0$ into the system (\ref{sysinf}) and taking $w
=0$, a two-dimensional system arises 

\begin{eqnarray} 
\frac{d\,u}{d\,\tau} &=& u\,\left( 2 +
\frac{v^2}{6} -\frac{u^2}{3} \right) \nonumber \\ 
\frac{d\,v}{d\,\tau} &=& v\,\left( -1 +
\frac{v^2}{6} -\frac{u^2}{3} \right) \;. 
\label{sysinf1}
\end{eqnarray} 

\noindent {}Five critical points are found: two repelling nodes $P_2 \;
(u=0,v=\sqrt{6})$, $P_2^{\prime} \; (u=0,v=-\sqrt{6})$, one saddle $P_0
\; (u=0,v=0)$, and $P_1 \; (u=\sqrt{6},v=0)$, $P_1^{\prime} \;
(u=-\sqrt{6},v=0)$ which are degenerate. In Fig. 3, these points are
shown on the plane $w = 0$ together with the unphysical trajectories on
it. Note that the projection of the cone $\rho_{rad}=0$ onto the plane
$w = 0$ is an integral curve of the system (\ref{sysinf1}). The regions
near $P_1$, $P_1^{\prime}$ are the projections of the inflationary
domain $2\,u^2 - v^2 \geq 0$ on $w = 0$. According to the phase portrait
shown in Fig. 3, the inflationary domain is an attractor for most
trajectories. Actually this is a general feature of the system, as can
be seen from Fig. 4, where we have plotted $\frac{\ddot{a}}{a}$ as a
function of time. Initial values were purposefully taken away from the
inflationary region. Note how the system is driven to the inflationary
domain in the different cases of dissipation. 

Curves emanate from $P_2$ and $P_2^{\prime}$ to the interior
of the cone. The asymptotic behavior near these points is 

\begin{equation} 
H = \frac{1}{3\,t},\;\;\; \phi =
\pm\,\sqrt{\frac{3}{4\,\pi}}\,M_{\rm pl}\,{\rm ln}\,(m\,t) +
\frac{v_0\,M_{\rm pl}}{2\,\sqrt{8\,\pi}}\,(m\,t)^{\frac{2}{3}} \mp
\frac{\beta_0}{m\,M_{\rm pl}}\,t, 
\end{equation} 

\noindent where $v_0$ is an arbitrary constant and the signs $\pm$
indicate which critical point $P_2$, $P^{\prime}_2$ we are considering.
The emergence from these points corresponds to the growth of time $t$
from the instant $t=0$ of the initial singularity. Near these
singularities, $\dot{\phi}^2 \gg \phi^2$, meaning that the kinetic
energy of the scalar field is dominant over the potential $V(\phi)$
producing an effective equation of state $p_{\phi} = \rho_{\phi}$.
Therefore all curves emanating from $P_2$, $P_2^{\prime}$ are not in the
inflationary regime. Note that the influence of dissipation starts to be
considerable as far as the trajectories evolve away from the critical
points. It is also possible to integrate the dynamical system near the
critical point $P_0$, whose asymptotic behavior near it is characterized
by 

\begin{equation} 
H = \frac{1}{2\,t},\;\;\; \phi =
\pm\,\frac{M_{\rm pl}}{2\,\sqrt{\pi\,m\,t}}, 
\end{equation} 

\noindent 
where the initial singularity is denoted by $t = 0$. 

The points $P_1$, $P_1^{\prime}$ lie on the inflationary region and are
attractors of trajectories on the invariant manifold $w=0$, suggesting
that they are also attractors for trajectories on the physical region
but close to the invariant manifold. It can be shown that there is only
one trajectory emerging from $P_1$ and $P_1^{\prime}$ into the physical
region, whose asymptotic solution is 

\begin{equation} 
H = -\frac{1}{3}\,m\,t,\;\;\; \phi =
\pm\,\frac{m\,M_{\rm pl}}{\sqrt{8\,\pi}}\,t, 
\label{asysol1} 
\end{equation} 

\noindent where $t$ increases from $-\infty$, the initial singularity.
As expected, this solution is inflationary, since $m^2\,\phi^2 \gg
\dot{\phi}^2$, which produces an effective equation of state
$p_{\phi}=-\rho_{\phi}$. All remaining trajectories that visit a small
neighborhood of these points experience an inflationary phase as
described above. It is worth pointing out that the asymptotic solution
(\ref{asysol1}) corresponds to the direction of approach given by
$v=-\frac{\sqrt{6}}{3}\,w$, which is equivalent to no dissipation (see
analysis carried out by Belinskii et al\cite{belins2}). The influence of
dissipation, from a purely dynamical system analysis, appears as a
second order effect changing the former relation to
$v=-\frac{\sqrt{6}}{3}\,w + \frac{\sqrt{6}}{9}\,\beta_0\,w^2$. 

\subsection{Case n=2} 

The structure of the phase space for $n=2$ is quite different from the
previous case. After substituting $n=2$ into the system (\ref{sysinf}),
we obtain a two-dimensional dynamical system whose phase portrait is
shown in Fig. 5. There are two lines of degenerate critical points,
denoted by $l_1 \; (-\infty < u < \infty,v=w=0)$ and $l_2 \; 
(-\infty < v < \infty,u=w=0)$. 
Here the projection of the cone surface $\rho_{rad}=0$,
represented by the dashed circle on Fig. 5, is not an integral curve of
the system at infinity. By continuity arguments, taking into account the
unphysical trajectories on $w=0$, we infer that the trajectories within
the physical region, but near the plane $w=0$, originate outside the
region $\rho_{rad}=0$, eventually outside the quantum domain, i.e.,
where the classical description is valid. Therefore, they must be ruled
out of our analysis. Nonetheless, some critical points on the lines
$l_1$, $l_2$ may be source of trajectories in the physical region. In
fact, using standard methods to analyze sets of degenerated critical
points \cite{bogoy}, we come to the conclusion that the points at
infinity $P_0 \; (u=v=0)$, $P_1 \; (u=\sqrt{6},v=0)$,
$P_1^{\prime} \; (u=-\sqrt{6},v=0)$, $P_2 \; (u=0,v=\sqrt{6})$,
$P_2^{\prime} \; (u=0,v=-\sqrt{6})$ are the relevant ones and belong to a
distinct nature from the remaining point, as we are going to describe.
From the points $P_0$, $P_2$ and $P_2^{\prime}$ three-dimensional
pencils of trajectories emerge, whose asymptotic behavior near the
points $P_0$ and $P_2$, $P_2^{\prime}$, are given, respectively, by 

\begin{equation} 
H = \frac{1}{2\,t},\;\;\; \phi =
\pm\,\frac{M_{\rm pl}}{2\,\sqrt{\pi\,m\,t}}, 
\end{equation} 

\begin{equation} 
H = \frac{1}{3\,t},\;\;\; \phi =
\pm\,\sqrt{\frac{3}{4\,\pi}}\,M_{\rm pl}\,{\rm ln}\,(m\,t). 
\end{equation} 

\noindent 
In the above solutions the initial singularity is
characterized by $t=0$, or equivalently the points $P_0$ and $P_2$,
$P_2^{\prime}$, and the emergence from the singularity corresponds to
increase of time from this value. We remark that the solutions are not
inflationary, but eventually evolve to this stage. 

The most interesting class of solutions emanate from the points $P_1$,
$P_1^{\prime}$ on the line $l_1$ (see Fig. 5). It can be shown that a
two-dimensional pencil of paths emerges from these points along the same
plane of the line $l_1$ and orthogonal to the plane $w=0$. The
asymptotic behavior in this region is characterized by $v = 0$, which
implies that the form of the solutions is the same as found in
(\ref{asysol1}). 

\section{Advantageous Trajectories and Radiation in \break Inflation
with Dissipation} 

We now have a closer look at those trajectories in phase space that may
lead to a long enough inflationary stage and let us try to quantify
these trajectories in a way similar to that in \cite{belins1}. During an
inflationary period given by $\Delta t = t_f - t_i$, the scale factor
changes by 

\begin{equation} 
\frac{a(t_f)}{a(t_i)} = \exp \left( \int_{t_i}^{t_f}
H(t) dt \right) \;. 
\label{a/a} 
\end{equation} 

\noindent 
Let us determine the above ratio in terms of the initial and
final values for the scalar field, $\phi_i$ and $\phi_f$, respectively. 

First note that the inflationary epoch can be characterized by those
trajectories that satisfy 

\begin{equation} 
m \phi \gg | \dot{\phi} | \;. 
\label{ir} 
\end{equation}

\noindent 
We consider the slow-roll approximation which, in the presence
of dissipation, can be written as 

\begin{equation} 
\dot{\phi} \simeq - \frac{m^2 \phi}{3 H + \eta (\phi)}
\;. 
\label{slow-row} 
\end{equation} 

\noindent 
We also assume that $\rho_\phi \gg \rho_{\rm rad}$ (see
\cite{Fang} and \cite{berera2}). Then, from (\ref{G1}) and (\ref{ir}),
we can write 

\begin{equation} 
H \simeq \sqrt{\frac{4 \pi}{3}} \frac{m |\phi|}{M_{\rm pl}}
\;. 
\label{H} 
\end{equation} 

Using (\ref{H}) and (\ref{slow-row}) in (\ref{a/a}), we get the explicit
expression: 

\begin{equation} 
\frac{a(t_f)}{a(t_i)} \simeq \exp \left[\frac{2
\pi}{M_{\rm pl}^2} \left( \phi_i^2 - \phi_f^2 \right) - \sqrt{\frac{4 \pi}{3}}
\int_{\phi_i}^{\phi_f} \frac{\eta(\phi)}{m M_{\rm pl}} d \phi \right] \;.
\label{a/a-eta} 
\end{equation} 

\noindent 
{}For $\eta (\phi) =0$ we recover the usual expression for the
scale factor ratio \cite{revinflat,belins1}. Taking $\eta (\phi) =
\eta_n \phi^n$, $n=0,\; 2$, we finally obtain 

\begin{equation} 
\frac{a(t_f)}{a(t_i)} \simeq \exp \left[\frac{2
\pi}{M_{\rm pl}^2} \left( \phi_i^2 - \phi_f^2 \right) + \sqrt{\frac{4 \pi}{3}}
\frac{\eta_n}{m M_{\rm pl}} \frac{(|\phi_i|^{n+1} - |\phi_f|^{n+1})}{n+1}
\right] \;. 
\label{a/a-eta_n} 
\end{equation} 

\noindent 
The term in the above exponential gives the total number, $N$,
of e-folds during inflation. We have numerically checked the validity of
the above equation. The physically interesting trajectories are as usual
those with $N \gtrsim 70$ \cite{revinflat}. 

With no dissipation, $\eta_n =0$, $N \sim 70$ can be accomplished with
$\phi_i \approx 3.4 M_{\rm pl}$ and $\phi_f \approx 0.2 M_{\rm pl}$
\cite{revinflat}. In the presence of dissipation, for a fixed $N$,
dissipation decreases the values of both $\phi_i$ and $\phi_f$. {}For
the values of $\phi_i$ and $\phi_f$, which $N\sim 70$ with $\eta_n =0$,
we see from (\ref{a/a-eta_n}) that dissipation begins to change
considerably $N$, say by more than $10 \%$, for $\eta_0 \gtrsim m$, or
$\eta_2 \gtrsim 0.25 m/M_{\rm pl}^2$ (in terms of the dimensionless
factors $\beta_n$, $\beta_0 \gtrsim 1$ and $\beta_2 \gtrsim 0.01$,
respectively). In both cases we have an initial dissipation of order
$\eta(\phi) \gtrsim m $. Since $H \approx m$, we must have $H \lesssim
\eta(\phi)$ in order for dissipation to begin changing the inflationary
phase considerably. 

The authors in \cite{belins1,belins2} have characterized the measure of
disadvantageous trajectories ($N < 70$) as $\sim m/M_{\rm pl}$, 
obtained by the
ratio of the trajectories beginning at $|\phi| < \phi_i$, in relation to
the complete quantum boundary, of length $2 \pi M_{\rm pl}^2/m$. 
Since for fixed
$N$ the effect of dissipation is to reduce $\phi_i$, we find that
in the presence of
dissipation $m/M_{\rm pl}$ ($\sim 10^{-7}$) is the upper bound for 
the measure
of disadvantageous trajectories. Dissipation turns inflation even more
general. 

We can also evaluate the energy of the inflaton field transferred to
radiation during the inflationary stage, from, for example, Eq.
(\ref{drad}). In Fig. 6 we plot $\rho_{\rm rad}$ against time for four
different values of $\beta_n$. The mark indicates when the trajectories
leave the inflationary region ($\ddot{a}/a$ becomes negative). We find
that there is a value of $\beta_n$ for which radiation is maximal, at
the exit of the inflationary stage, given by $\beta_0$, $\beta_2 \approx
1$ ($\eta_0 \approx m$ and $\eta_2 \approx 25 m/M_{\rm pl}^2$). These
results are fairly independent of the initial conditions. Also, we can
find an ideal range for $\beta_n$ such that we exit inflation smoothly
to the radiation era with the Universe sufficiently hot. At
thermalization 

\begin{equation} 
\rho_{\rm rad} = \frac{\pi^2 g_*}{30} T^4 \;,
\label{radT} 
\end{equation} 

\noindent 
where $g_*$ ($\sim 10^2 - 10^3$) is the effective number of
degrees of freedom. If we impose that at the end of inflation we should
have at least $T \sim m$ ($m\simeq 5.10^{-7} M_{\rm pl}$ for the
quadratic potential model \cite{revinflat}), we have numerically
obtained, for both $\beta_0$ and $\beta_2$, the approximate bound: 

\begin{equation} 
10^{-7} \lesssim \beta_n \lesssim 800 \;. 
\label{bound}
\end{equation} 

\noindent 
Too high values of $\beta_n$ lead to a longer inflationary
period. Most of the inflaton's energy is transferred to radiation which
is red-shifted away before the end of inflation. There is no 
possibility of a
reheating phase for these cases. On the other hand, very small values of
$\beta_n$ can provide a smooth exit from inflation to radiation and the
inflaton field energy density can still provide an additional 
reheating phase by the usual mechanisms\footnote{At least for $\beta_0 < 2$,
when we are in the underdamped regime (see Sec. II) and for any value of
$\beta_2$ smaller than the upper value in (\ref{bound}).}. The maximum
temperature reached by the thermal bath, at the exit of the inflationary
stage in our model, is $T\approx 10^{15} GeV$, for $\beta_0, \; \beta_2
\approx 1$. 

\section{Discussions and Conclusions} 

We have examined a dynamical system describing inflaton dynamics in
the presence of dissipation. Dissipation is taken as being present throughout
the inflaton's evolution and we have presented the changes in the
phase diagram due to the effect of dissipation. Our results allow us to
draw a series of important conclusions concerning the dynamical system
and the physical system itself. 

We have shown that the continuous production of radiation due to dissipation
changes effectively the inflaton dynamical system. We have distinguished two
main consequences for the field trajectories in phase space due to the
presence of dissipation: (a) In field independent dissipation (for $\beta_0
\geq 2$) the trajectories tend towards the $z$-axis at the final stages of
evolution. In this regime the inflaton's motion is overdamped. The stronger
the dissipation ($\beta_0 \gg 2$) the faster the field trajectories are
attracted to the $z$-axis; (b) In field dependent dissipation the motion is
initially damped but the field trajectories always tend to the cone surface
$\rho_{\rm rad} = 0$. The stronger the dissipation the slower the field
trajectories tend to this surface. 

{}From the physical system side we can also reach some interesting conclusions
which may have important consequences for inflationary dynamics. We have seen
that dissipation associated with inflaton decay not only supports inflation
but also draws all inflationary field trajectories into a longer stage of
inflation. This is true for both kinds of dissipation we have considered. In
terms of a measure of the amount of trajectories possessing the required
inflationary period, a similar analysis to the one given in
\cite{belins1,belins2} allows us to reach the conclusion that inflation is
even more general in the presence of dissipation. 

Our results also allow us to reach some conclusions concerning warm
inflation scenarios or non-isentropic inflation, in particular concerning
the problem of reheating. In non-isentropic inflation, as opposed to
standard inflation, there is the possibility of no reheating phase,
due to the continuous production of entropy and heat, leading to a
smooth exit from inflation to the radiation era \cite{berera2,nonisen}.
We then expect that in these scenarios there is a strong damping of the
inflaton field's motion in the direction of the minimum of the
inflaton's potential. In this paper we have quantified how strong the
dissipation must be, when compared to the friction due to expansion, in order
to achieve favorable scenarios of radiation at the exit from the
inflationary stage. We have shown that no reheating is only achieved in
the overdamped regime ($\beta_0 > 2$) for constant dissipation, when
there is no oscillatory motion of the inflaton around the potential
minimum, or for very high values of $\beta_2$, when all the inflaton's
energy is transfered to radiation and this is then red-shifted away due
to the longer expansion. 

A further question we may ask is about a possible realization of chaotic
behavior in Cosmology. According with some previous in the realm of
isotropic and anisotropic Bianchi IX cosmologies\cite{chaos}, the
Universe can experience two possible outcomes: collapse into the big
crunch after an initial stage of expansion and expansion into the
inflationary regime. In this way, chaotic behavior is associated with an
indetermination of the final state (collapse or expansion) of the
Universe once a set of initial conditions in an infinitesimal region is
chosen. We may say that the boundaries of collapse and expansion are
mixed. In our case\footnote{Also similar conclusions are reached for the
model studied by Belinskii et.al. \cite{belins1,belins2}.} all orbits
initially in the physical region $\rho_{rad} \geq 0$ are unavoidably
attracted to the state of unlimited expansion represented by the origin
of the phase space (cf. Fig. 1). The orbits can not randomically explore
the region inside the cone $\rho_{rad} = 0$, since the volume of the
phase space is not conserved. Therefore, we can conclude that chaotic
behavior does not take place for those orbits lying in the physical
region in the cases studied previously, i. e., for $n=0, 2$. Such
conclusions are also valid for the case analyzed by Belinskii and
Khalatnikov \cite{belins2} in the realm of anisotropic models. 

{}Finally one of the most important results of this paper was to show
that even a value of dissipation as small as $\eta(\phi) \sim 10^{-7} m$
(compare with $H \sim m$) can still lead to very important consequences
for inflationary dynamics, such as providing a smooth exit from inflation to
radiation, and it also shows that the interaction of the inflaton field
with other degrees of freedom, manifested in the form of dissipation,
cannot in general be neglected. We are currently working on a
detailed microscopic model for dissipation in inflation, to be presented
in a forthcoming paper. 

\vspace{0.7cm} 

\begin{center} 
{\large \bf Acknowledgements} 
\end{center}

\vspace{1.0cm}

We would like to thank D. Boyanovsky and A. Berera for useful
discussions. We also thank J. E. Skea for reading and 
helping us to revise the text.
This work was partially supported by Conselho Nacional de
Desenvolvimento Cient\'{\i}fico e Tecnol\'ogico - CNPq (Brazil).

\newpage 

\centerline{\large \bf Figure Captions} 

\vspace{2cm} 

\noindent 
{\bf Figure 1:} The field trajectories in phase space
(projected in the plane $x,z$) for the cases: (a) no dissipation; (b)
constant dissipation ($\beta_0 = 0.3$ for dashed line, $\beta_0 = 0.7$
for full line) and (c) field dependent dissipation ($\beta_2 = 10$ for
full line and $\beta_2 = 90$ for dashed line). Initial conditions for
both cases are $(x_0,y_0,z_0)=(\sqrt{6}, 10^{-3}, 2.5)$. Dotted lines
represent the lines where $\rho_{\rm rad} =0$.

\vspace{0.25cm}

\noindent 
{\bf Figure 2:} View of the physical region of the phase
space. The infinity is compactified onto the plane $w=0$. 

\vspace{0.25cm}

\noindent 
{\bf Figure 3:} Phase portrait of the plane $w=0$ for the case
of constant dissipation ($n=0$). The inflationary domain is represented 
as the projection into the
plane $w=0$. 

\vspace{0.25cm}

\noindent 
{\bf Figure 4:} Plot of $\frac{\ddot{a}}{a}$ against time, in
the units of (\ref{sys2}), for the cases of: no dissipation (full line);
constant dissipation ($\beta_0=10$), dashed line; and field dependent
dissipation ($\beta_2=10/6$), dotted line. Initial conditions for both
cases are $(x_0,y_0,z_0)=(\sqrt{6},10^{-3},2.5)$. 

\vspace{0.25cm}

\noindent 
{\bf Figure 5:} Integral curves at infinity for the case of
field dependent dissipation. The dashed circle represents the projection
of the cone $\rho_{rad}=0$. The projection
of the inflationary domain, into the plane $w=0$, is the same as in Fig. 3. 

\vspace{0.25cm}

\noindent 
{\bf Figure 6:} $\rho_{rad}$ for different values of
dissipation. The mark indicates the exit of the inflationary stage.
Initial conditions are
$(\phi_i,\dot{\phi_i},H_i)=(3.4\,M_{\rm pl},0,3.4\,\sqrt{\frac{4\,\pi}{3}}\,
m)$\, ($\rho_{rad_i}=0$).

\newpage

\epsfysize=10cm
\centerline{\epsfbox{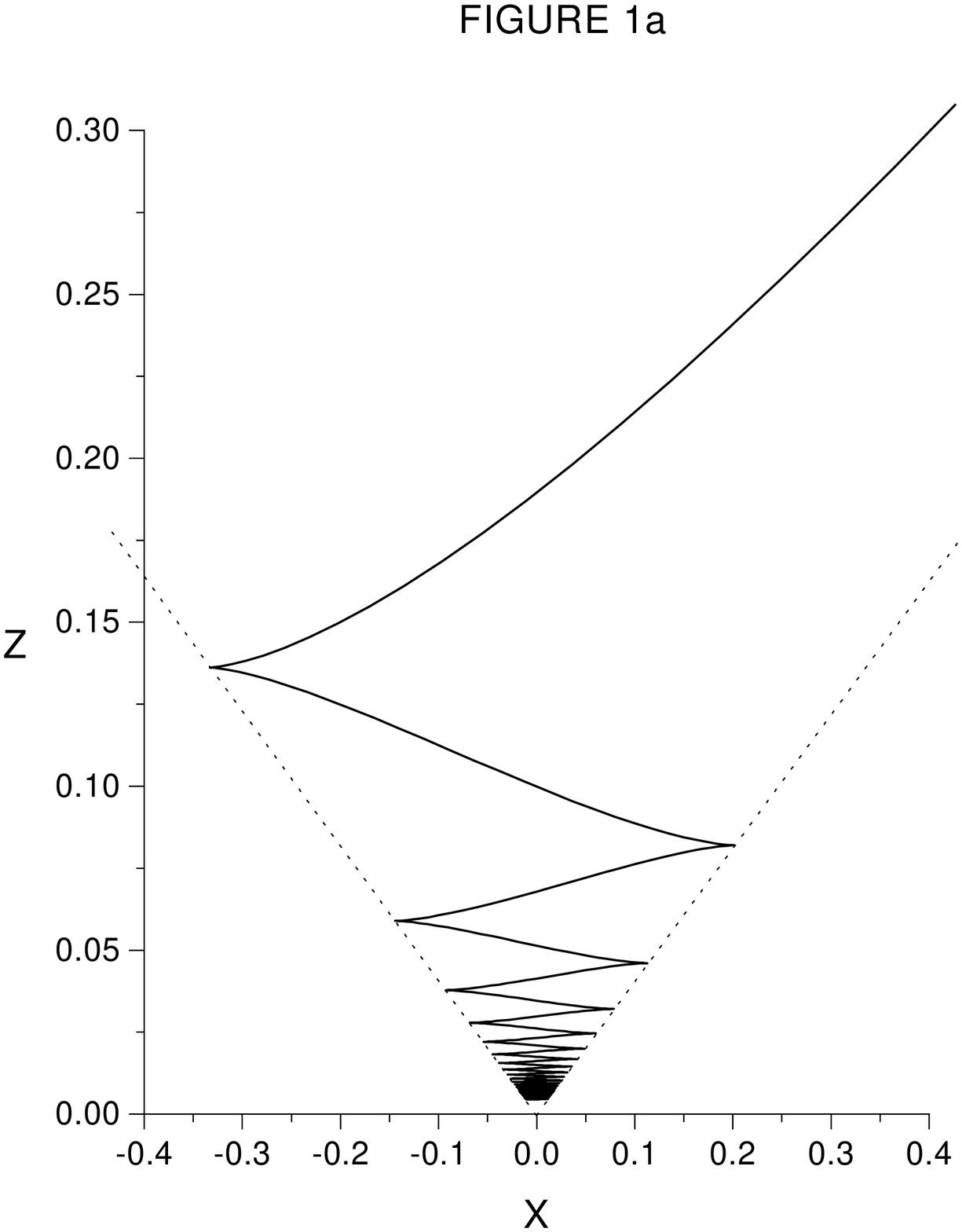}}

\epsfysize=10cm
\epsfbox{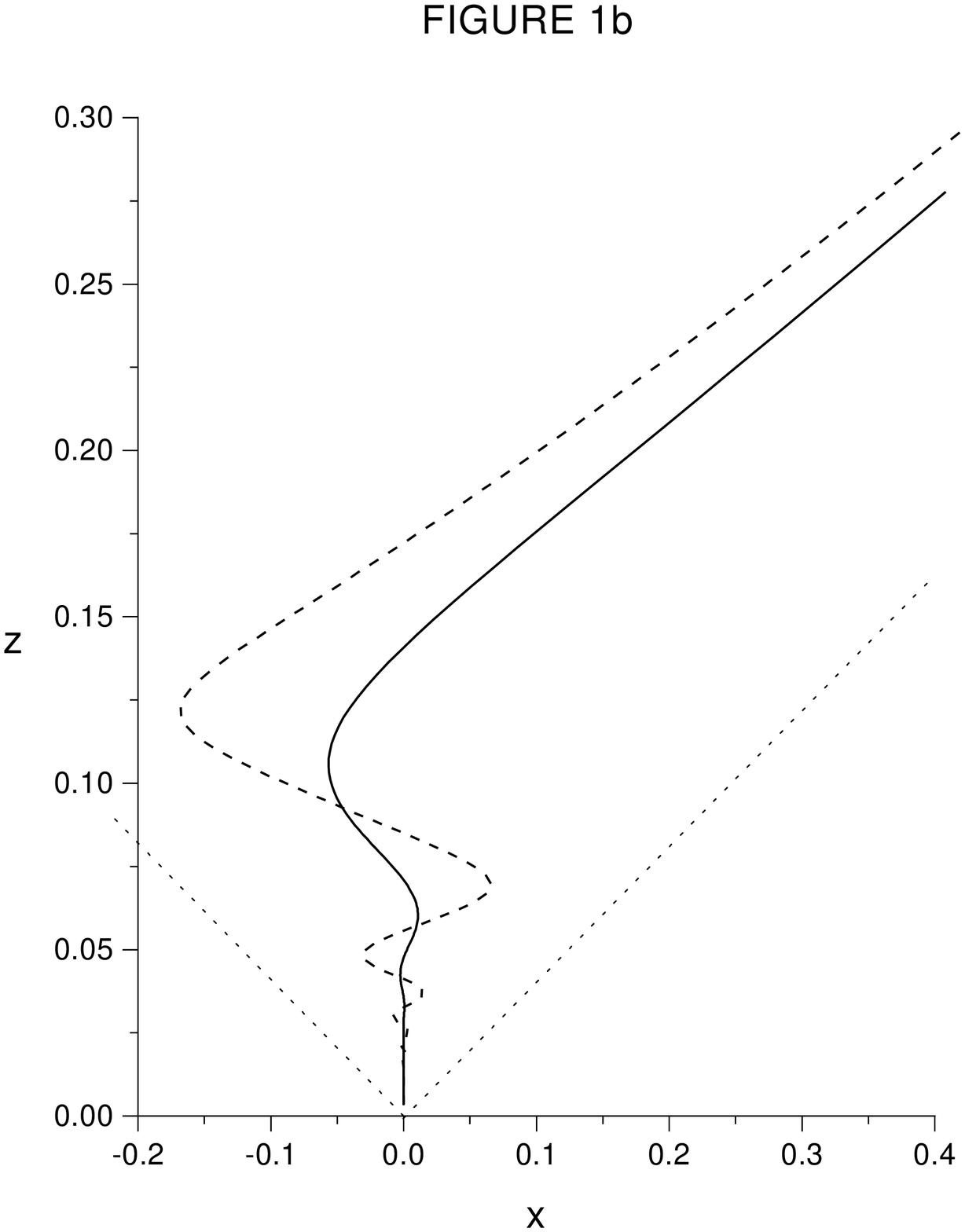} 

\vspace{-10cm}

\epsfysize=10cm
\hfill{\epsfbox{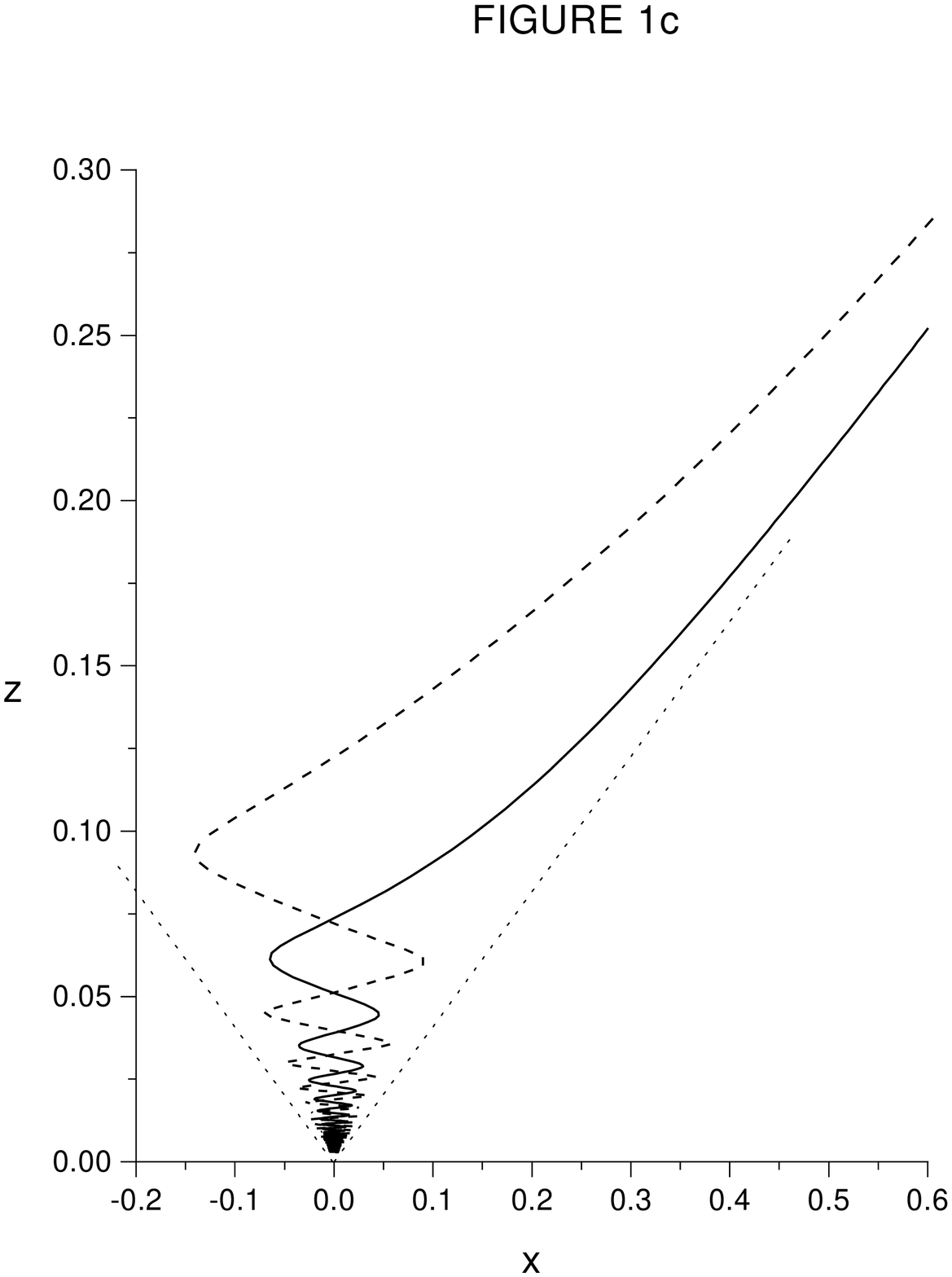}}

\newpage

\epsfysize=8cm
\centerline{\epsfbox{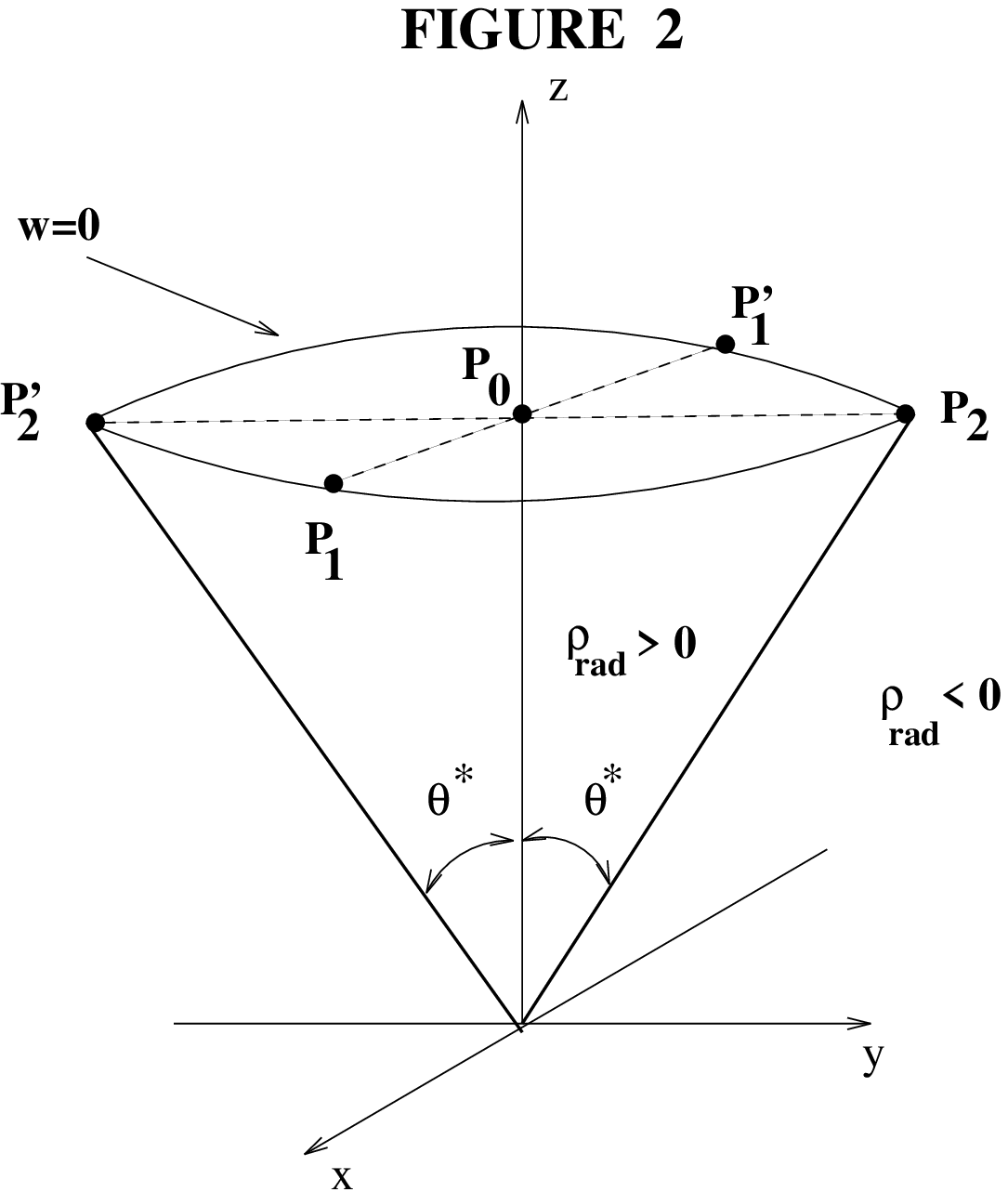}}

\vspace{1.6cm}

\epsfysize=8cm
\centerline{\epsfbox{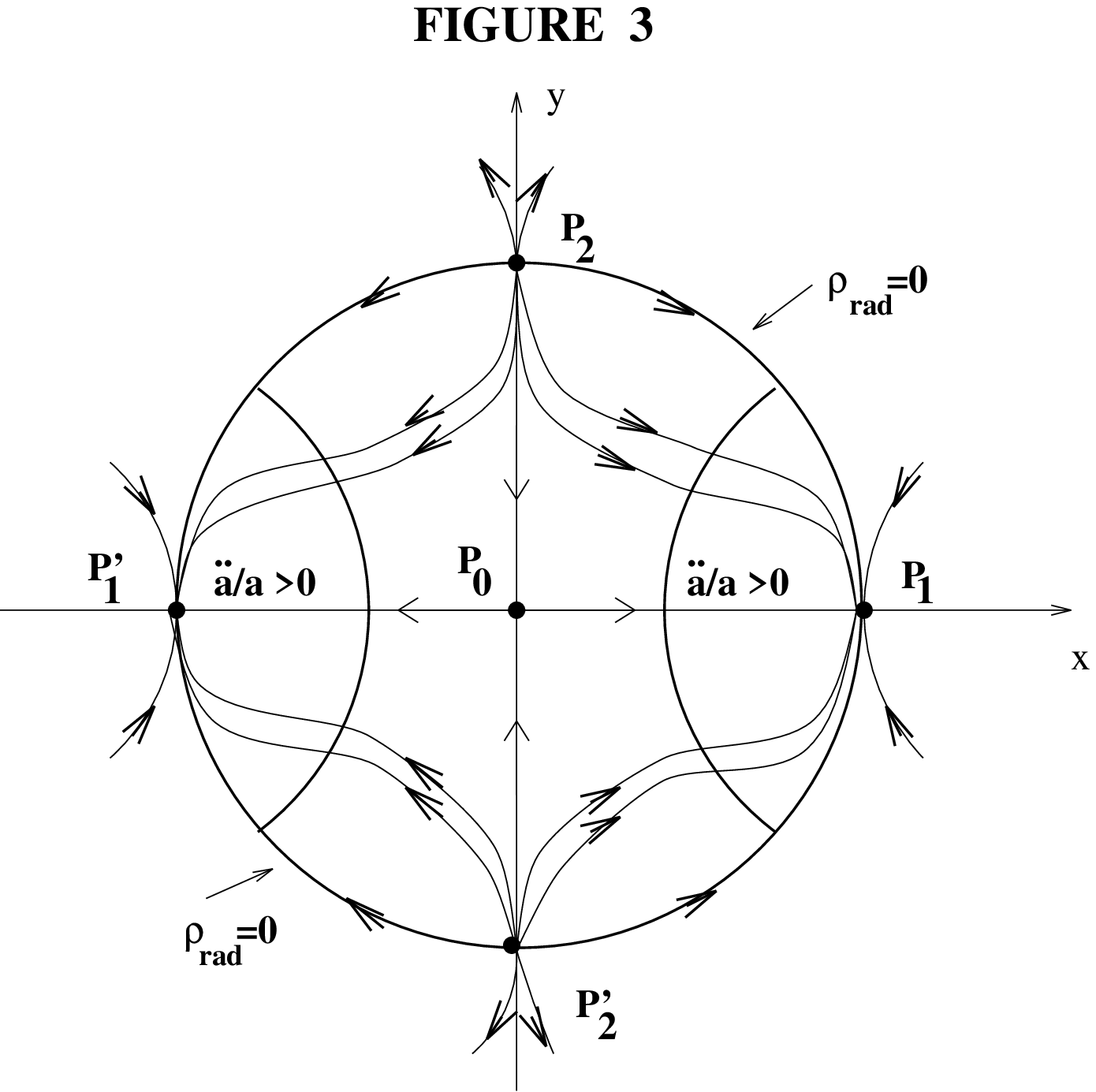}}

\newpage

\vspace{-3.7cm}

\epsfysize=13cm
\centerline{\epsfbox{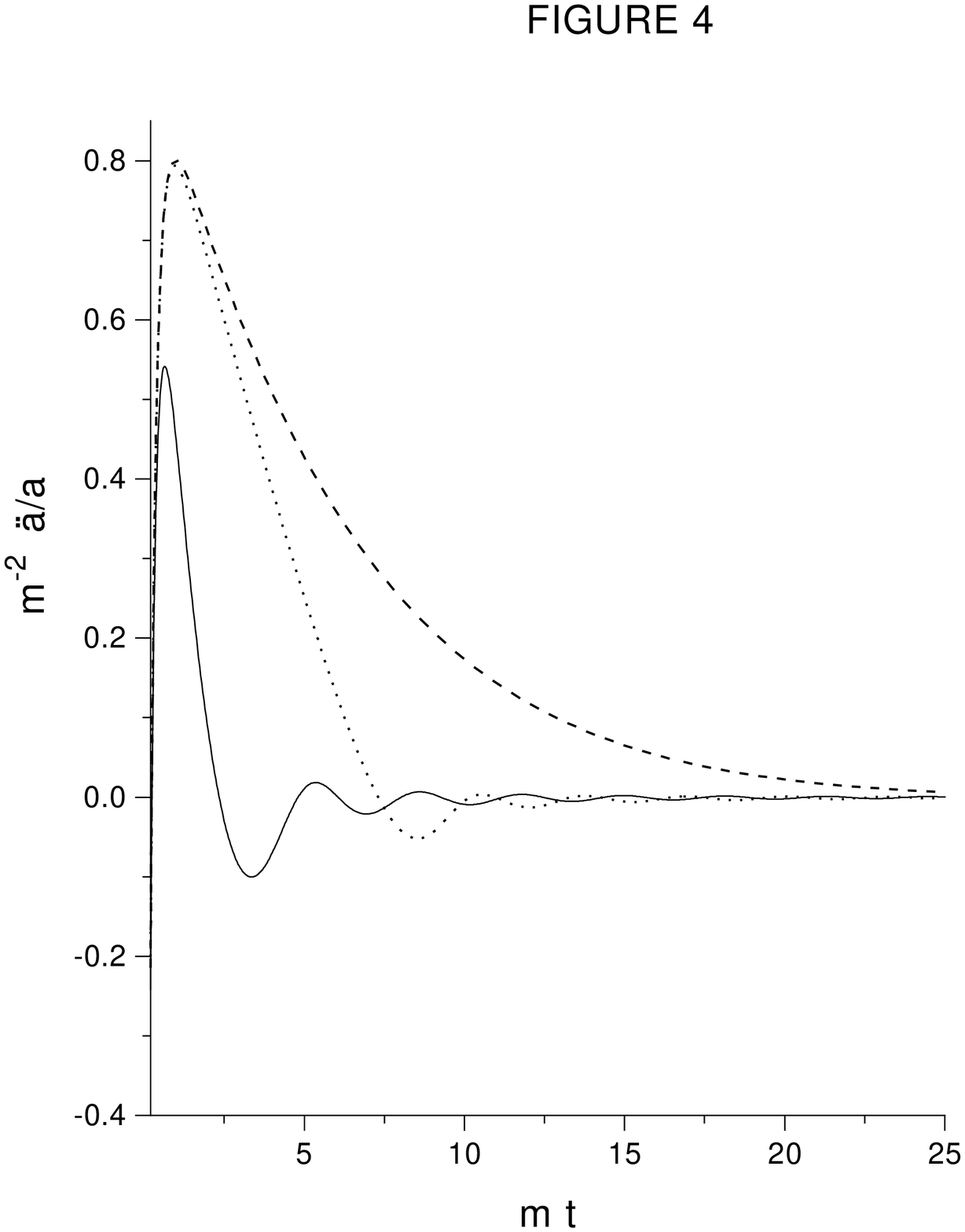}}

\vspace{0.5cm}

\epsfysize=7cm
\centerline{\epsfbox{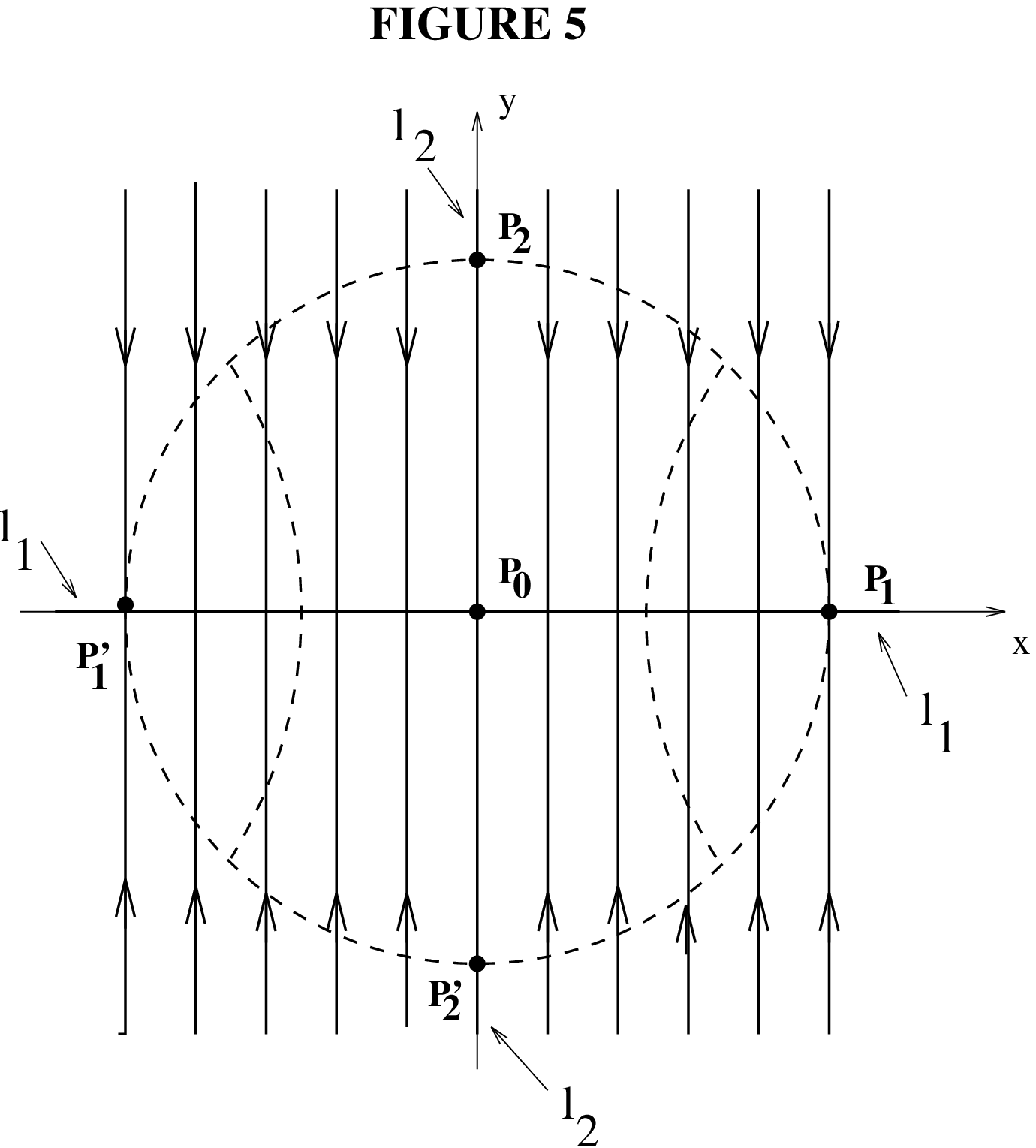}}

\newpage

\epsfysize=16cm
\centerline{\epsfbox{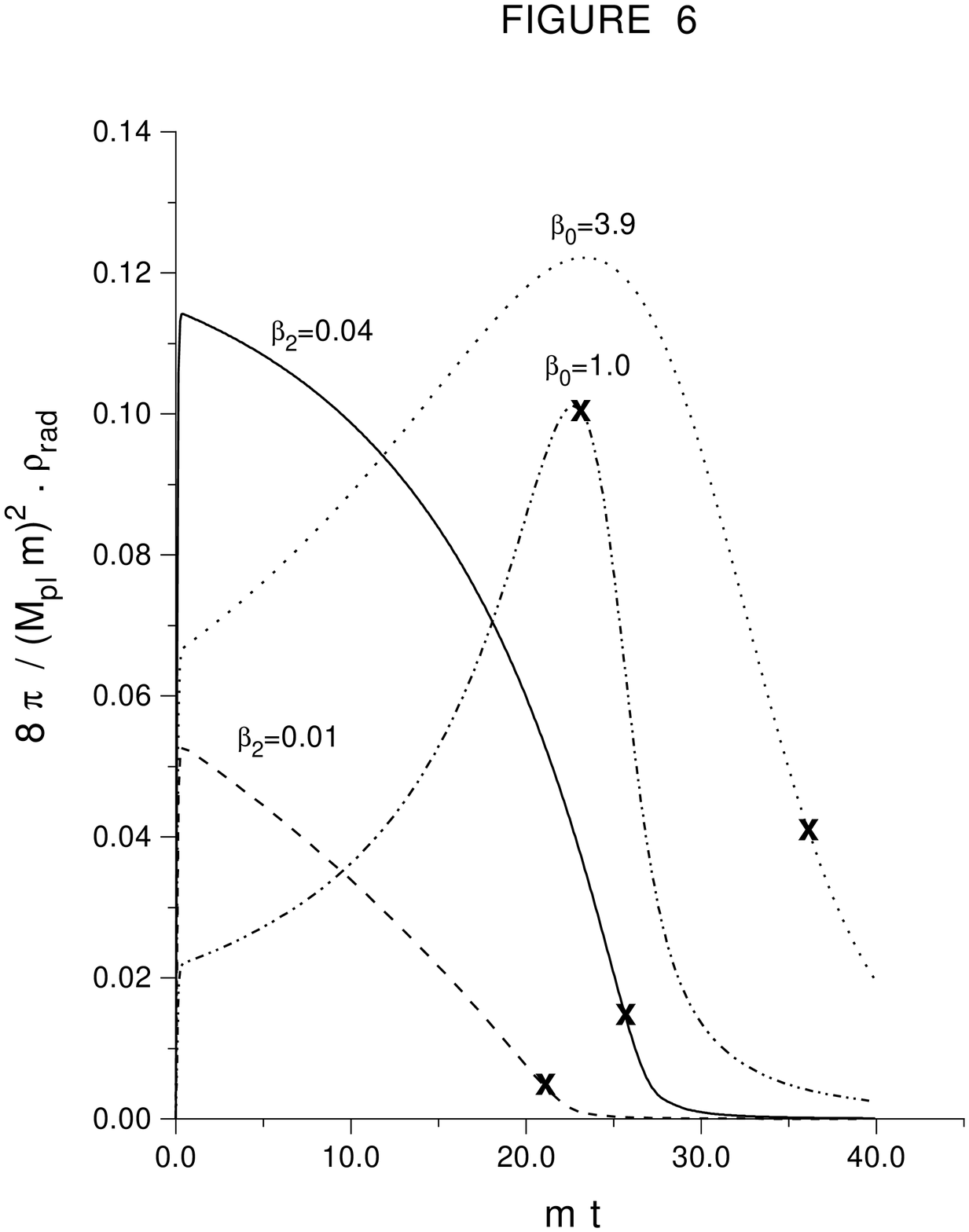}}

\end{document}